\documentclass[epj]{svjour}
\usepackage{latexsym}
\usepackage{graphics}
\usepackage{amssymb}
\begin{document}
\title{Yukawa model on a lattice: two body states}
\author{F. De Soto\inst{1,2} \and J. Carbonell\inst{2} \and C. Roiesnel\inst{3}
Ph. Boucaud\inst{4} \and J.P. Leroy\inst{4} \and O. P\`ene\inst{4}
}                     
\offprints{F. de Soto, fcsotbor@upo.es}          
\institute{Departamento de Sistemas F\'{\i}sicos Qu\'{\i}micos y
Naturales. Universidad Pablo de Olavide, 41013 Sevilla (Spain)\and
Laboratoire de Physique Subatomique et Cosmologie 53 av. des
Martyrs, 38026 Grenoble (France)\and Centre de Physique
Th\'eorique Ecole Polytechnique, UMR7644, 91128 Palaiseau,
(France) \and Laboratoire de Physique Th\'eorique, UMR8627,
Universit\'e Paris-XI, 91405 Orsay (France) }
\date{Received: date / Revised version: date}
%
\abstract{We present first results of the solutions of the Yukawa
model as a Quantum Field Theory (QFT) solved non perturbatively
with the help of lattice calculations. In particular we will focus
on the possibility of binding two nucleons in the QFT, compared to
the non relativistic result.
} 
\maketitle
\section{Introduction}
\label{intro}

Nucleon-nucleon ($NN$) interaction is probably one of the most
studied problems in theoretical physics. From meson exchange
models~\cite{OBEM,Pieper} till effective chiral
la\-gran\-gians~\cite{chiral}, the effort of physicists has been
towards the development of suitable $NN$ potentials that, once
included in a Lipmann-Schwinger (LS) equation, would provide the
nuclear binding energies and scattering properties. Using Green's
Function Montecarlo one can even compute the nuclear spectrum of
nuclei up to $\sim$12 nucleons \cite{Pieper}.

Most potential models are inspired by an underlying Quantum Field
Theory (QFT), from which only a very particular kind of diagrams
are taken into account when solving the LS equations: in practice
the resolution is currently available only in the ladder
approximation.

All crossed-ladder graphs were summed up for the Wick-Cut\-kos\-ky
(WC) model~\cite{tjon}, and the resulting binding energies are
much bigger than those obtained within the ladder approximation.
This strong bias is one of the most important motivations for the
present work. As the WC model is not consistent as a field
theory~\cite{baym}, we will study the simplest renormalizable QFT
involving fermions, where one species of fermions interact with a
scalar meson via a Yukawa coupling.

The interest of this approach is manifold. On one hand it allows a
comparison with the results of the ladder approximation in
different relativistic and non relativistic equations. On the
other hand, and including other cou\-plings, it could provide a
relativistic description of nuclear ground states in terms of the
traditional degrees of freedom -- mesons and nucleons -- with no
other restriction than those arising from the structureless
character they are assumed to have.

\vspace*{-.2cm}
\section{The model}
\label{model}

We consider a system of two identical fermions ($\psi$)
interacting through the exchange of a scalar meson ($\phi$)
described by the lagrangian density,
\begin{equation}\label{lagrangian}
\mathcal{L}=\overline\psi D \psi + \mathcal{L}_{KG}(\phi) +
g_0\overline\psi\phi\psi\ ,
\end{equation}
where $D=\gamma_\mu\partial_\mu-M_0$ is the Dirac operator with a
bare fermion mass $M_0$, $\mathcal{L}_{KG}$ is a Klein-Gordon
lagrangian for the  scalar field. In the NR limit
(\ref{lagrangian}) gives rise to the potential
\begin{equation}
V(r)=-\frac{g_0^2}{4\pi} \frac{e^{-m_s r}}{r}\ ,
\end{equation}
where $m_s$ is the meson mass. The NR model does depends on a
unique parameter, $G=\frac{g_0^2}{4\pi} \frac{M}{m_s}$, and the
first bound state appears for $G\approx1.68$. The existence of
this unique scaling parameter $G$ can be easily shown in
Schrodinger equation but it is no longer true for the relativistic
case or the QFT.

In order to study the bound states, one needs to take into account
contributions to all orders in the coupling. Therefore, a
perturbative approach is not suitable. Instead, a genuinely non
perturbative tool will be used, lattice field theory, developed in
the context of QCD. The lattice Yukawa model is solved in a
Euclidean space-time where vacuum expectation values  are computed
in the Feynman path integral approach. For the dressed nucleon
propagator one has, for instance,
\begin{eqnarray}\label{correlator}
G^{\alpha\beta}(x,y)&=&\langle 0|\psi^\alpha(x)
\overline\psi^\beta(y)|0\rangle =\\
&=&\frac{1}{Z}\int[d\overline\psi][d\psi][d\phi]\psi^\alpha(x)
\overline\psi^\beta(y) e^{-S_E(\overline\psi,\psi,\phi)}\
,\nonumber
\end{eqnarray}
where the euclidean action acts as a probability distribution,
allowing for a Montecarlo integration.

We have chosen the following discretization of scalar fields:
\begin{equation}\label{kg}
S_{KG}=\frac{1}{2}\sum_x\left[\left(8+a^2m_s^2\right)\phi^2_x-2\sum_\mu\phi_{x+\mu}\phi_x\right]
\end{equation}
and  for fermion ones:
\[
S=\sum_{x y} \overline\psi_x D_{x y} \psi_y\ ,
\]
where $D_{xy}$ is the Wilson-Dirac operator:
\begin{eqnarray}\label{wilson}
D_{x y} &=& \left(1+g_L\phi_x\right)\delta_{x, y} - \nonumber
\\ &-& \kappa\sum_{\mu} \left[ (1-\gamma_\mu)\delta_{x+\hat\mu, y}
+ (1+\gamma_\mu)\delta_{x-\hat\mu, y} \right]\ .
\end{eqnarray}
in which the hopping parameter, $\kappa=1/(8+2aM_0)$ and
$g_L=2\kappa g_0$ have been introduced.

Fermion fields -- being Grassmann variables -- have to be
integrated out in an algebraic way, resulting into:
\begin{equation}\label{prop}
G^{\alpha\beta}(x,y)=\frac{1}{Z}\int[d\phi] {D^{-1}}^{\alpha
\beta}_{x y} \det(D) e^{-S_{KG}}\ .
\end{equation}
This calculation is rather demanding in computing time due to the
determinant. The task is considerably  simplified in the
``quenched'' approximation, which consists in neglecting all
virtual nucleon-antinucleon pairs originated from the meson field
$\phi\to\bar\psi\psi$. Because of the heaviness of the nucleon,
this appears as a good approximation for the problem at hand and
has been adopted all along this work. Note that this is not a
priori justified for QCD, where quarks are very light.
Nevertheless, the quenched approximation gives there qualitatively
good results. Mathematically it is equivalent to set $\det(D)=1$.
The main numerical task in calculating (\ref{prop}) is the
inversion of Dirac operator $D_{xy}$

In the quenched approximation, and in absence of meson
self-interaction terms, the meson fields are free, and $\phi$
field configurations can be independently generated by a gaussian
probability distribution in momentum space.

\vspace*{-.2cm}
\section{Spectrum of Dirac operator}

\begin{figure}
\resizebox{0.475\textwidth}{!}{%
  \includegraphics{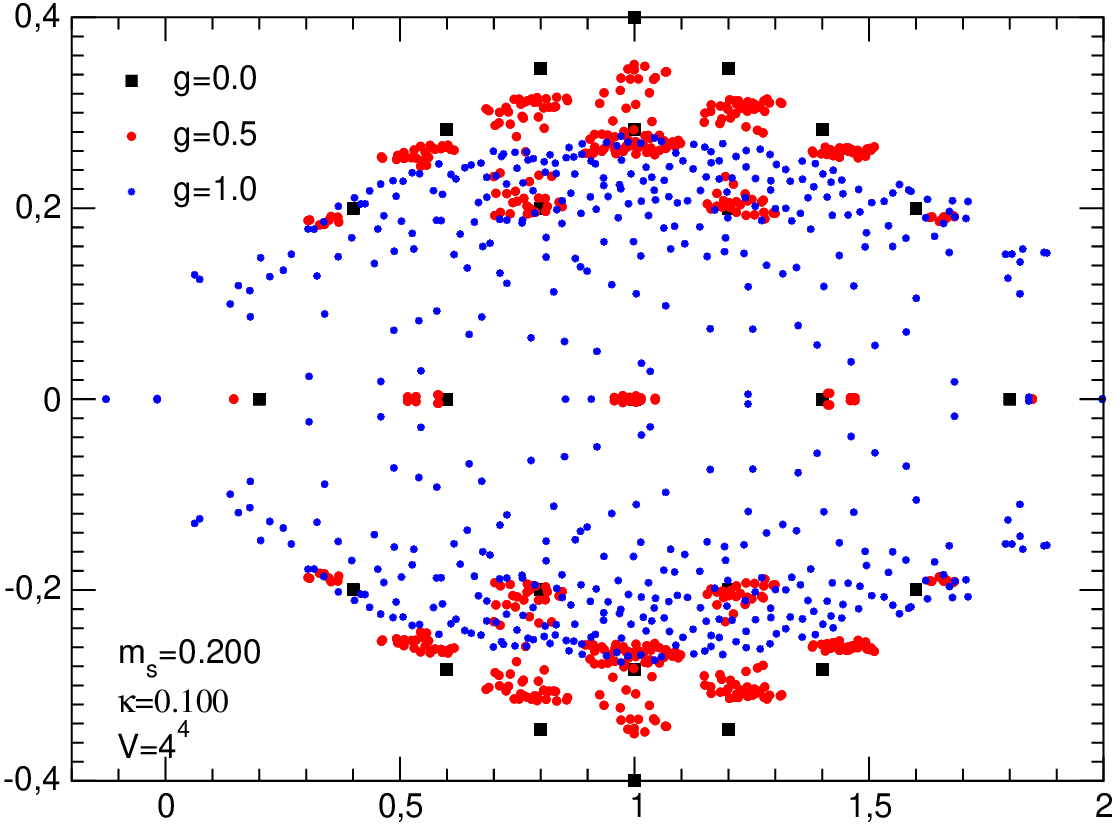}
}
\caption{Spectrum of Wilson-Dirac operator for a small lattice and
several values of the coupling.}
\label{fig:spectre}       
\end{figure}

The spectrum of the Dirac operator (\ref{wilson}) in the free case
lies in a circle centered in $\lambda=(1,0)$ and with radius
$8\kappa$. In QCD when the interaction is tuned up, the
eigenvalues are modified, but its real part is always bounded from
below. In the Yukawa model, on the contrary, the coupling term
plays the role of a mass: as the coupling constant grows, the
spectrum spreads out in real part and some eigenvalues go to the
negative real part half-plane (figure \ref{fig:spectre}).

\begin{figure}
\resizebox{0.475\textwidth}{!}{%
  \includegraphics{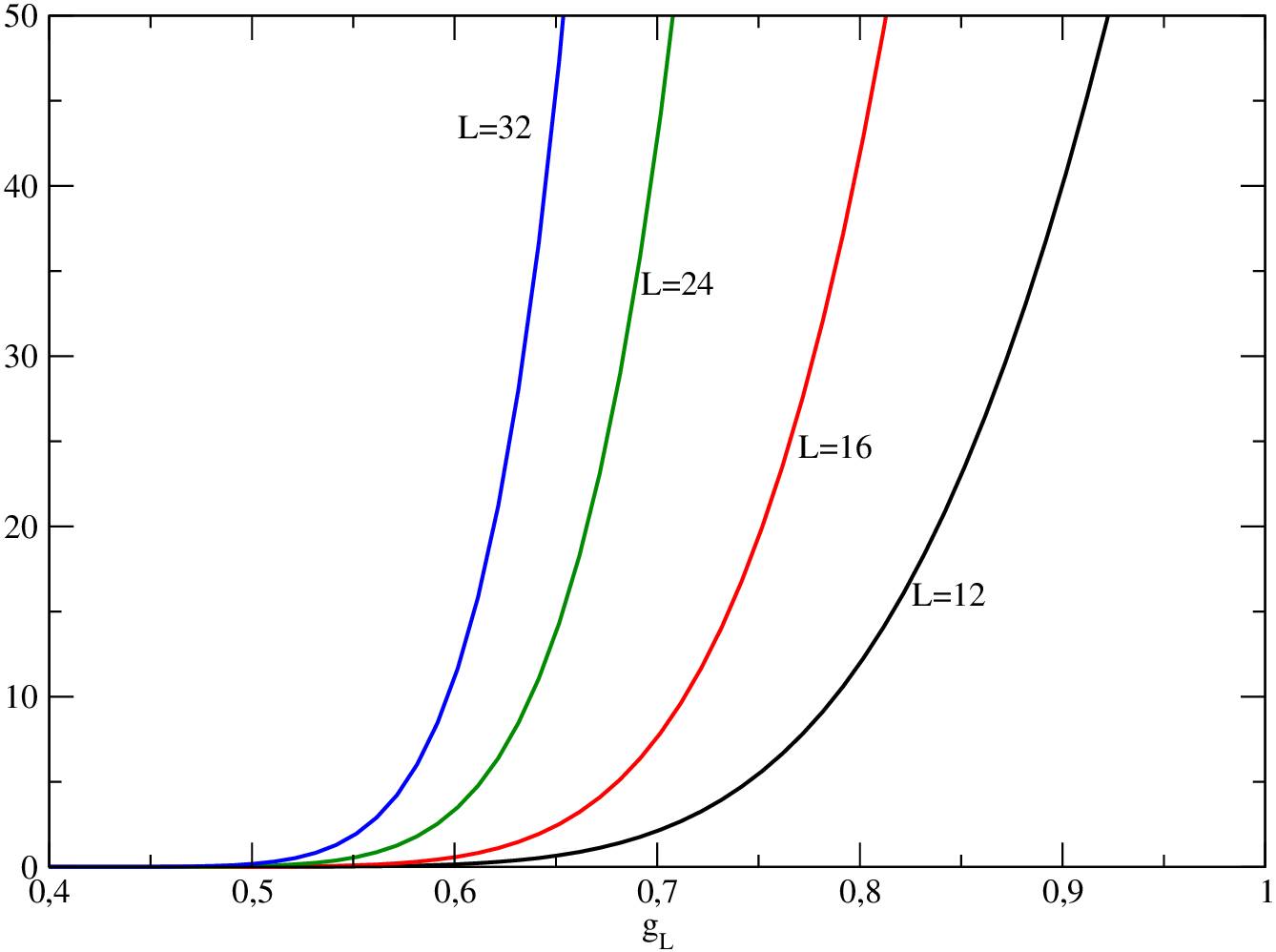}
}
\vspace*{0.4cm} \caption{Average number of negative eigenvalues
for several lattices with $L a m_s=5$ as a function of
$g_L$.}\vspace*{-0.5cm}
\label{fig:prob}       
\end{figure}

Negative real part eigenvalues spoil the convergence of most
iterative algorithms, but is not a fundamental problem. For larger
values of the coupling and large lattices, nevertheless, the
probability of having one -or more- eigenvalues very small grows
dramatically. A simplified but significant picture can help to
estimate the appearance of those small eigenvalues. The diagonal
terms in (\ref{wilson}) have the form $1+g_L\phi_x$, that will be
zero as soon as $\phi_x=-1/g_L$. The values of $\phi_x$ are
distributed according to (\ref{kg}), as a gaussian of width
\begin{equation}
\sigma^2 = \sum_{k\in V} \frac{1}{\hat k^2 + m_s^2}\ ,\qquad
\hat{k}^2=2\sum_\mu(1-\cos(k_\mu)).
\end{equation}
The probability of having one negative eigenvalue is plotted in
figure \ref{fig:prob} for different lattice sizes, showing how
negative eigenvalues appear for $g_L\sim0.5$. The non diagonal
terms in (\ref{wilson}) modify this picture, and in practice there
are small eigenvalues for $g_L\gtrsim0.8$, hindering the numerical
solution of the linear system. This implies that there is a
maximum value of the coupling constant that can be used in this
model. The problem could perhaps be solved in the unquenched case,
as the fermionic determinant would eliminate the configurations
with very small $\det(D)$.

\vspace*{-.2cm}
\section{One and two-body masses}

One-body mass is computed from the time dependence of euclidean
correlators:
\begin{equation}
C_1(t)=\sum_{\vec{x}} Tr\left[G(\vec{x},t)\right]\sim e^{-M_1 t}\
,
\end{equation}
for large values of $t$, determining the fermion renormalized
mass, $M_1$. Preliminary results on one body masses for both
scalar and pseudoscalar coupling were already presented in
\cite{qcd05}.

Two body masses, $M_2$, are obtained in a similar way, from the
time evolution of the propagator of an operator
$J(x)=\Gamma_{\alpha\beta} \psi_\alpha(x) \psi_\beta(x)$ creating
a nucleon pair,
\begin{equation}\nonumber
C_2(t) = \sum_{\bf x} Tr \langle J({\bf x},t)J^\dagger({\bf
0},0)\rangle\sim e^{-M_2 t}\ ,
\end{equation}
that for large values of $t$ projects on the lowest energy state
with the quantum numbers of the operator $J(x)$. The matrix
$\Gamma$ determines the spin and parity of the state, being
$\Gamma=i\gamma_2\gamma_0\gamma_5$ for $J^\pi=0^+$ (ground state)
and $\Gamma=\gamma_2\gamma_0$ for $J^\pi=0^-$.

\begin{figure}
\resizebox{0.475\textwidth}{!}{%
  \includegraphics{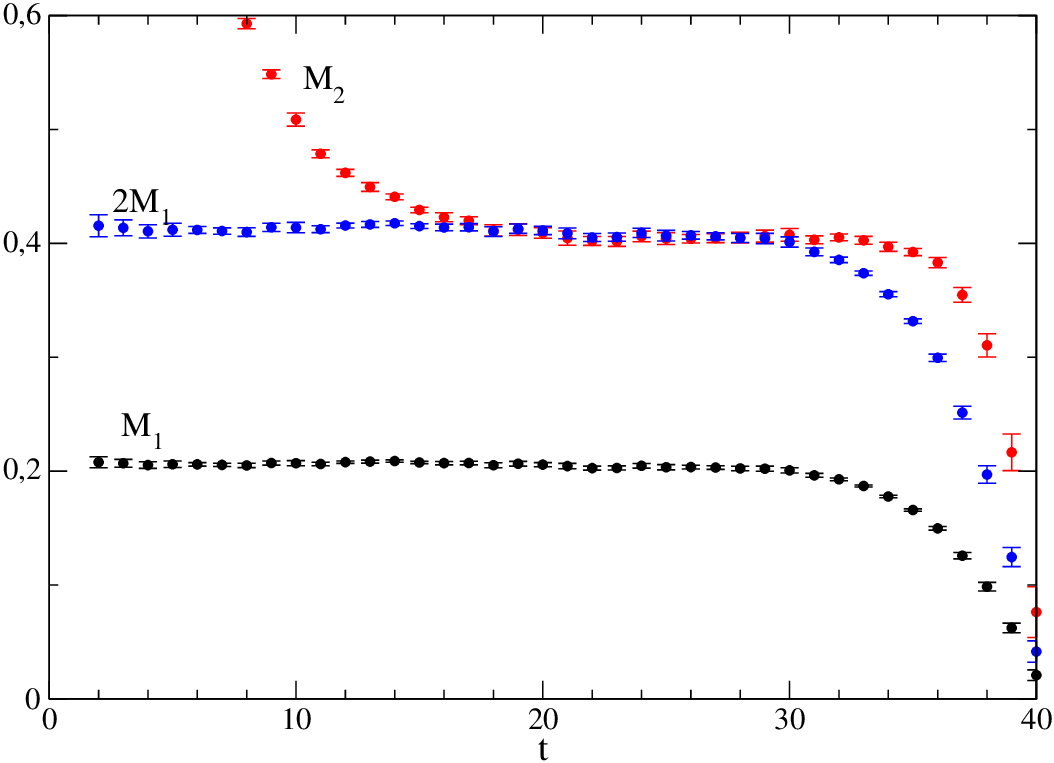}
}
\caption{One and two body effective masses as a function of the
euclidean time, for a $20^3\times80$ lattice with $g_0\approx1.6$
and $am_s=0.200$. In blue two times fermion mass, to compare with
mass of two fermion $0^+$ state.}
\label{fig:meff}       
\end{figure}

The exponential behavior is reached only at large values of $t$.
It is useful to define an effective mass as:
\[
M_{eff}(t)=\ln\left(\frac{C(t)}{C(t+1)}\right)
\]
that tends for large $t$ to the mass of the state and helps to
find the adequate fitting window. Some results can be found in
figure \ref{fig:meff} for one and two body masses.

In the figure twice fermion mass is plotted for comparison. With
these parameters twice fermion mass is not distinguishable of two
fermion mass. This is the common picture for the whole set of
parameters tested, no signal of the existence of a bound state is
found below the critical value of the coupling constant~\footnote{
There might exist bound states for very light mesons, near the
Coulomb limit, but this regime is difficult to reach on the
lattice due to the hierarchy of scales appearing.}.

\vspace*{-.2cm}
\section{Discussion}

Renormalization effects have been analised for one body masses,
where perturbation theory works, and the renormalization issues
concerning the coupling constant have also been
discussed~\cite{qcd05}.

The existence of a maximum value of the coupling in a QFT
treatment of the Yukawa model has been established. This critical
value is smaller than the one needed to form a bound state in the
NR limit, and no signal of such a bound state for lower couplings
has been observed. This limit on the coupling is characteristic of
the QFT, different from the potential approach where the coupling
usually take values $G\gg 1$.

    The meaning of this result needs to be clarified. It may be related
to the quenched approximation, or the fact that we neglect meson
self interaction. But then it should be noted that the same
approximations are performed in the non relativistic (Schrodinger)
treatment.
    For a given value of the lattice spacing, there exists other
ways to discretize the nucleon-meson interaction which don't have
these zero modes. This has to be further studied and it is not
clear if it allows to reach larger renormalized coupling constants
and particularly to reach the bound regime. We are not yet in a
position to decide if the bound on the coupling constant we
encounter is a lattice artefact or if it really casts a doubt on
the Yukawa theory itself.

It is known that the Yukawa theory is infrared free and, as such,
encounters the ``triviality problem'' i.e. that the ultraviolet
cut-off can not be driven to infinity without the theory becoming
trivial. It means that this can only be an effective theory with a
physical ultraviolet cut-off. The problem encountered from the
difficulty to invert the Dirac operator seems also to put a
limitation on the continuum limit. It is not clear whether both
problems are related or just happen both to hinder the continuum
limit.

Whether we manage or not to overcome the difficulty of reaching
the domain where bound states appear, the connection between the
QFT treatment and the Schrodinger approach can still be performed
by an estimate of the scattering  parameters which can be computed
thanks to the method proposed by Luscher~\cite{Luscher}, and have
recently been reexamined  in~\cite{ours,fb18}. 

%

\end{document}